\documentclass[12pt]{article}
\usepackage{amsmath,amssymb,exscale}
\usepackage{graphicx}
\usepackage{epsfig}
\usepackage{multicol}
\usepackage{color}
\usepackage{mathrsfs }
\usepackage{blindtext}
 \usepackage{fancyhdr}
\usepackage{hyperref}
\usepackage{cite}

\numberwithin{equation}{section}

\usepackage{xcolor}
\usepackage{sectsty}

\usepackage{mdframed}
\usepackage{titletoc}

\definecolor{secnum}{RGB}{13,151,225}
\definecolor{ptcbackground}{RGB}{212,237,252}
\definecolor{ptctitle}{RGB}{0,177,235}

\titlecontents{lsection}
  [5.8em]{\sffamily}
  {\color{secnum}\contentslabel{2.3em}\normalcolor}{}
  {\titlerule*[1000pc]{.}\contentspage\\\hspace*{-5.8em}\vspace*{5pt}%
    \color{white}\rule{\dimexpr\textwidth-15.5pt\relax}{1pt}}

\usepackage{hyperref}
\hypersetup{colorlinks,bookmarksopen,bookmarksnumbered,citecolor=rossos,
linkcolor=redy,pdfstartview=FitH,urlcolor=green-go}
\usepackage{slashed}

\definecolor{blus}{cmyk}{1,1,0,0.1}
\definecolor{verdes}{cmyk}{0.99,0,0.59,0.65}
\definecolor{rossos}{cmyk}{0,1,1,0.55}
\definecolor{redy}{cmyk}{0,1,1,0.7}
\definecolor{greeny}{cmyk}{0.99,0,0.59,0.98}
\definecolor{green-go}{cmyk}{0.79,0,0.59,0.5}

\usepackage{titlesec}

\def\hhref#1{\href{http://arxiv.org/abs/#1}{arXiv:#1}} 

\newcommand{\tmtextbf}[1]{{\bfseries{#1}}}
\newcommand{\tmtextrm}[1]{{\rmfamily{#1}}}

\def\be{\begin{equation}}
\def\ee{\end{equation}}
\def\ba{\begin{array} }
\def\bac{\begin{array} {c}}
\def\bacc{\begin{array} {cc}}
\def\baccc{\begin{array} {ccc}}
\def\bacccc{\begin{array} {cccc}}
\def\ea{\end{array}}
\def\bea{\begin{eqnarray}}
\def\eea{\end{eqnarray}}

\definecolor{red}{rgb}{1,0,0}

\def\psl{\hbox{\hbox{${p}$}}\kern-1.9mm{\hbox{${/}$}}}
\def\dsl{\hbox{\hbox{${\partial}$}}\kern-2.2mm{\hbox{${/}$}}}
\def\Dsl{\hbox{\hbox{${D}$}}\kern-2.6mm{\hbox{${/}$}}}
\newcommand{\bp}{\bar M_{\rm Pl}}

\newcommand{\gappeq}{{\rlap{{\raise}.5ex\text{\ensuremath{>}}}{{\lower}.5ex\text{\ensuremath{\sim}}}}}
\newcommand{\lappeq}{{\rlap{{\raise}.5ex\text{\ensuremath{<}}}{{\lower}.5ex\text{\ensuremath{\sim}}}}}

\newcommand{\I}{\tmtextrm{1{\kern}-.24em l}}

\begin{document}
\topmargin -1.0cm
\oddsidemargin -0.5cm
\evensidemargin -0.5cm

\vspace{-1cm}
 \hspace{12cm}IFT-UAM/CSIC-15-140
 
{\vspace{1cm}}
\begin{center}
\vspace{1cm}

 {\huge  \tmtextbf{ 
{\color{blus} Higgs Stability and the 750 GeV Diphoton Excess  }}} {\vspace{.5cm}}\\
%

\vspace{1.4cm}

{\large  {\bf Alberto Salvio$^{a,b}$, Anupam Mazumdar$^{c}$
}
\vspace{.3cm}

{\it }\

{\em  \normalsize 

$^a$~Theoretical Physics Department, CERN, Geneva, Switzerland\\

\vspace{0.2cm}

$^{b}$~Departamento de F\'isica Te\'orica, Universidad Aut\'onoma de Madrid\\ and Instituto de F\'isica Te\'orica IFT-UAM/CSIC,  Madrid, Spain \\

\vspace{0.2cm}

$^{c}$~Consortium for Fundamental Physics, Lancaster University, Lancaster, LA1 4YB, UK}

\vspace{0.5cm}

}
\vspace{1.9cm}
 \end{center}
\noindent --------------------------------------------------------------------------------------------------------------------------------

\vspace{-0.3cm}

\begin{center}
{\bf  Abstract}
\end{center}

\noindent {\normalsize    We study the implications of a possible unstable particle with mass $M_X<$ TeV for the Higgs stability, naturalness  and inflation. We pay particular attention to the case $M_X\approx$ 750 GeV, suggested by recent results of ATLAS and CMS on diphoton final states, and work within the minimal model:  we add to the Standard Model  field content a pseudoscalar and a vector-like fermion carrying both color and electric charge. This can stabilize the   electroweak  vacuum without invoking new physics at very high energies, which would give an unnaturally large contribution to the Higgs mass. We also show that inflation can be obtained via a UV modification of General Relativity.   }


\vspace{.9cm}

\noindent --------------------------------------------------------------------------------------------------------------------------------

\vspace{1.1cm}


\newpage

{\color{redy}
\tableofcontents
}

\vspace{.9cm}
\noindent --------------------------------------------------------------------------------------------------------------------------------

\section{Introduction} \label{introduction}

The experiments at the LHC have recently released the first studies at the highest energy scale ever reached in a collider, $\sqrt{s}=13$ TeV. Besides confirming the Standard Model (SM) predictions in many observables, the ATLAS~\cite{ATLAS} and CMS~\cite{CMS} collaborations have reported an excess in the diphoton channel at $750$ GeV, with a local significance of $3.6\sigma$ and $2.6\sigma$, respectively, possibly due to a resonance, whose width $\Gamma$ might be relatively large: ATLAS results suggest $\Gamma/M_X \approx 0.06$. However, the statistical preference over a narrow width is very small \cite{Falkowski:2015swt}; indeed, CMS suggests a narrow width. If a large width, $\Gamma/M_X \approx 0.06$, is assumed for CMS, then the significance decreases to $2.0\sigma$. Note that this excess of di-photons at $750$ GeV is not accompanied by any missing energy, leptons, i.e. $\ell^{+}\ell^{-}$, $ZZ$,  or jets, nor that the $\sqrt{s}= 8$ TeV data showed any anomaly apart from a slight upward fluctuation at $m_{\gamma\gamma}\approx 750$ GeV.

Taking into account the Landau-Yang theorem~\cite{Landau,Yang}, this signal could be due to a boson with spin different from 1,  decaying into 2 photons. 
Of course, the simplest option would be to assume the presence of a new scalar field $X$, a singlet under the SM gauge group with a 
mass $M_X \approx 750$~GeV. The presence of any new boson, which can couple to photons, would naturally beg plethora of questions relevant for beyond the Standard Model physics and cosmology. One of the important issues is the
stability of the electroweak (EW) vacuum; within the SM the tussle between the top Yukawa  and the Higgs self-coupling suggests a metastable vacuum for the Higgs, which has been confirmed at the 2.8$\sigma$ level by computing two-loop corrections in Ref.~\cite{Buttazzo:2013uya,Degrassi:2012ry}. Indeed, the metastability of the EW vacuum can cause severe restrictions on the possible models of the early universe \cite{Espinosa:2015qea}.
There are also other cosmological questions if one introduces a new scalar field: how would $X$ couple to dark matter~\cite{Mambrini:2015wyu}? Or
 could $X$ play the role of the inflaton, or the curvaton?

The aim of this paper is to understand how our picture of the early universe would respond to the introduction of a new SM gauge singlet 
scalar field with mass below $1$ TeV, and decaying into photons at a high enough rate, if possible.  As usual, scalar fields
play a crucial role in  inflation, acting as secondary fields whose 
perturbations can potentially seed large scale structures, or being responsible for reheating the universe, or 
creating matter-anti-matter asymmetry (for a review see \cite{Mazumdar:2010sa}). 

In the simplest scenario, we wish to model the possible new spin-0 boson, $X$, as a real CP-odd scalar, which couples to
an extra vector-like fermion $\Psi$, carrying color and electric charge, and the SM Higgs\footnote{In this case the width is smaller than $\Gamma/M_X \approx 0.06$; we will ignore this possible issue here, given that such value has currently almost the same statistical significance as a narrow width.}. Furthermore, we will demand that the $X$ field does not develop a vacuum expectation value (VEV) at late times, 
otherwise it would break CP spontaneously. A natural question that we would like to answer is then the following: what would happen to the stability of the 
EW vacuum in the presence of another light scalar field $X$, which has couplings to the Higgs? Could the couplings 
be large enough, but still within perturbative limits, to explain the diphoton excess?
The second paramount question which we pose is: could this light $X$ be responsible for cosmic inflation like the SM Higgs with its non-minimal coupling to gravity?

Of course, there is the possibility that the diphoton excess will turn out to be a statistical fluctuation. Even in this case we believe our questions are interesting. One reason is that  if the EW vacuum is stabilized by  new physics that is not much above the TeV scale the Higgs mass is natural\footnote{See also   \cite{Salvio:2014soa,Kannike:2015apa} for an extension of these ideas to include gravity and inflation.}
     \cite{Farina:2013mla}, while if it is stabilized by much heavier physics a fine-tuning is required (modulo protection mechanisms such as supersymmetry, which so far have not been observed). 

In Section \ref{Minimal model} we introduce the model we will work with and discuss the typical values of the parameters that can account for the diphoton excess. 
 We will study the 
modified renormalization group equations (RGEs) up to 1-loop due to $X$ and $\Psi$ couplings and the matching conditions due to the new masses in Section \ref{RGE}. In fact, the diphoton excess would require a large Yukawa coupling between $X$ and $\Psi$, which contributes negatively to the self-coupling of $X$; this coupling, however,  is required
to be positive to maintain the stability of the EW vacuum. As we shall see in Section \ref{stab}, despite this fact, it is  possible to stabilize the EW vacuum. In Section \ref{stab} we will also  study the classical dynamics of gravity and the scalars, by considering the most general Lagrangian with operators of dimension up to 4, and investigate whether inflation can be realized.  Both $X$ and the SM Higgs can now couple to gravity, also
via non-minimal interactions, which we will take into account in our analysis. We provide our conclusions in Section \ref{Conclusions}.


\section{Minimal model} \label{Minimal model}

We consider the model with Lagrangian
\be \mathscr{L} =  \sqrt{-g}\left( \mathscr{L} _{\rm SM}+  \Delta \mathscr{L}+\mathscr{L} _{\rm gravity}\right),\label{full-lagrangian}
 \ee
 The gauge group is the SM one: 
  $G_{\rm SM}\equiv {\rm SU(3)_c\times SU(2)_{\it L}\times U(1)_{\it Y}}.$ 
Also, $g_{\mu\nu}$ is the metric of the space-time and $g$ its determinant.   $\mathscr{L} _{\rm SM}$ is the SM Lagrangian (minimally coupled to gravity). $ \Delta \mathscr{L}$ represents the beyond-the-SM terms in the Lagrangian due to the chosen  model.  We consider the following new physics.
\begin{itemize}
\item {\bf A Dirac fermion.} (In Weyl notation) it is a pair of two-component fermions $\Psi_1$ and $\Psi_2$ in the following vector-like representation of $G_{\rm SM}$: 
$ \Psi_1 \sim (3,1)_q, \,  \Psi_2 \sim (\bar{3}, 1)_{-q}.$
Namely, they form a Dirac fermion in the fundamental representation of $SU(3)_c$,   neutral under $\rm SU(2)_{\it L}$ and with electric charge $q e$ (which is left a priori as a free parameter).
\item {\bf A real pseudoscalar.} This scalar $X$, which may be identified  with a possible resonance at $\approx$ 750 GeV,  is  CP-odd and neutral under $G_{\rm SM}$.
\end{itemize}
Assuming renormalizability, the most general CP-even  $ \Delta \mathscr{L}$ is
\be  \Delta \mathscr{L} = i\sum_{j=1}^2\overline{\Psi}_j \Dsl \, \Psi_j -M_\psi(\Psi_1 \Psi_2 +\overline{\Psi_1 \Psi_2})+\frac12(\partial X)^2-\Delta V(H,X)  -iy\, X(\Psi_1 \Psi_2 -\overline{\Psi_1 \Psi_2}) + ...\label{DeltaL}  \ee
where $H$ is the Higgs doublet,
\be \Delta V(H,X) \equiv  \frac{m^2_X}{2} X^2+ \frac{\lambda_X}{4} X^4 + \frac{\lambda_{HX}}{2} ( |H|^2 -v^2/2) X^2 \ee
and the classical potential of the full model is
\be  V(H,X)=\lambda_H\left(|H|^2-v^2/2\right)^2+\Delta V(H,X). \ee
%

The dots in Eq. (\ref{DeltaL}) represent extra  Yukawa couplings and mixing terms between $\Psi$ and the  SM quarks, which are possible for  special values of the charge of $\Psi$; these couplings can make $\Psi$ unstable.
For example, for $q=2/3$  one can write down the gauge invariant operator $H\Psi_2 Q$, where $Q$ is the SM quark doublet. For simplicity  we take  these terms small enough
 that their contribution to the running of the relevant parameters is negligible\footnote{In Fig. \ref{RG-evolution} we will choose $q=3/2$, which forbids the above-mentioned operator.}. 

 Finally,  $\mathscr{L} _{\rm gravity}$    includes the pure gravitational Lagrangian and the possible non-minimal couplings between gravity and the other fields,  we consider
\be\mathscr{L} _{\rm gravity}=  - \frac{\bp^2}{2} R - \Lambda -  \left(\xi_H |H|^2 +\xi_X X^2/2\right) R + \alpha R^2,\ee
 where $\bp$ is the reduced Planck mass, $\Lambda$ is a cosmological constant, $\alpha$, $\xi_H$ and $\xi_X$ are real couplings and $ R$ is the Ricci scalar. We demand $\alpha$ to be positive, which ensures the stability of Minkowski's space-time. This $\mathscr{L} _{\rm gravity}$ is the most general gravity Lagrangian with parameters of dimension of non-negative powers of energy modulo the Gauss-Bonnet and the Weyl squared terms. The Gauss-Bonnet term can be written as a linear combination of the $R^2$ and Weyl squared term plus a total derivative, which does not affect the following analysis. The Weyl squared term, being invariant under conformal transformations and vanishing on the flat space-time, does not change the classical treatment of inflation. We will assume that the same is true at quantum level. 

Note that, even if we set $\alpha=0$, $\xi_H=0$ and/or $\xi_X=0$, the corresponding operators ($R^2$, $|H|^2R$ and  $X^2 R$) are generated by quantum corrections. Therefore, one should not regard the absence of such terms as a satisfactory option.

The requirement that $\Delta \mathscr{L}$ be CP-even simplifies considerably the model, avoiding terms that are odd under $X\rightarrow -X$ in the potential.\footnote{CP is broken by small SM Yukawa couplings: the effect of this breaking on $V$ is negligibly small for our analysis as suppressed by loop factors and the small value of such couplings.} This guarantees that the mixing between the Higgs and $X$ is small as suggested by data \cite{Falkowski:2015swt}. We introduce a colored fermion $\Psi$ because it easily ensures (through its Yukawa coupling $y$) that the production of $X$ is sizable in proton  collisions at the LHC.

In addition to the SM parameters and $q$, this model has   only 5 parameters if we ignore the couplings to gravity: $M_X$, $M_\psi$,  $\lambda_{HX}$, $\lambda_X$ and $y$. We take all of them to be real and positive.\footnote{At least at the classical level; whether they remain positive after quantum corrections  will be addressed later on.} 

The EW symmetry breaking is triggered by the VEV $v\approx 246\,$GeV of the neutral component  of the Higgs doublet.  We do not want to break CP spontaneously, therefore we take the VEV of $X$ to be zero.   Given that we assume $\lambda_{HX}>0$ and of course
\be \lambda_H >0 , \qquad \lambda_X>0,\label{stability condition} \ee
 for these values of the scalar fields we reach\footnote{This is the case because 
\be   M_X^2-\frac{ v^2\lambda_{HX}}{2} >0,\ee
 for a large $X$ squared mass, e.g. $M_X \approx 750$ GeV, and $\lambda_{HX}$ in the perturbative regime.} the absolute minimum of $V$. 


This model has been studied  in  \cite{Franceschini:2015kwy} (see also  \cite{Ellis:2015oso} and \cite{Falkowski:2015swt}). The partial decay rates for $X\rightarrow gg$ and   $X\rightarrow \gamma\gamma$  are 
\be \Gamma_{gg}\equiv \Gamma(X\rightarrow gg) = M_X \frac{\alpha_3^2}{8\pi^3} \tau y^2 |{\cal P}(\tau)|^2, \qquad \Gamma_{\gamma\gamma}\equiv\Gamma(X\rightarrow  \gamma\gamma) = M_X \frac{9 \alpha^2}{16\pi^3} q^4 \tau y^2 |{\cal P}(\tau)|^2 \ee
where $\tau\equiv  4 M_\psi^2/M_X^2$  and ${\cal P}(\tau)\equiv \arctan^2(1/\sqrt{\tau-1}))$. The reported excess  can be achieved for  $ \Gamma_{gg}/M_X\sim 10^{-3}-10^{-6}$ and  $ \Gamma_{\gamma\gamma}/M_X\sim 10^{-6}$.  Note that   $y, q \sim 1$ and $M_\psi \sim$ TeV can account for the claimed excess. 

%

\section{RGE analysis and thresholds} \label{RGE}

Since we want to study the predictions of this model at energies much above the EW scale, up to the Planck scale, we need the complete set of RGEs.  We use the modified minimal subtraction  ($\overline{\rm MS}$) renormalization scheme  to define the renormalized couplings and their RGEs. Moreover, for a generic renormalized coupling $g$ we write the RGEs as
\be \bar{\mu}^2\frac{dg}{d\bar{\mu}^2}= \beta_{g},\ee
where $\bar{\mu}$ is the $\overline{\rm MS}$ renormalization energy scale. The $\beta$-functions  $\beta_{g}$ can also be expanded in loops:
\be  \beta_{g} =  \frac{\beta_{g}^{(1)}}{(4\pi)^2}+ \frac{\beta_{g}^{(2)}}{(4\pi)^4}+ ... \, ,\ee 
where $ \beta_{g}^{(n)}/(4\pi)^{2n}$   is the $n$-loop contribution. 

Let us start from energies much above $M_X$ and $M_\psi$. In this case we have
\bea  \beta_{g_1^2}^{(1)}& =&   \frac{(24 q^2+41)g_1^4}{10}, \qquad   \beta_{g_2^2}^{(1)} =- \frac{19g_2^4}{6},\qquad\beta_{g_3^2}^{(1)}  = -\frac{19 g_3^4}{3},\nonumber\\   \beta_{y_t^2}^{(1)}  & =& y_t^2\left(\frac92 y_t^2-8g_3^2-\frac{9g_2^2}{4}-\frac{17g_1^2}{20} \right),\nonumber\\ 
  \beta_{\lambda_H}^{(1)} & =&\left(12\lambda_H+6y_t^2-\frac{9g_1^2}{10}-\frac{9g_2^2}{2}\right)\lambda_H-\,3y_t^4 +\frac{9 g_2^4}{16}+\frac{27 g_1^4}{400}+\frac{9 g_2^2 g_1^2}{40}+\frac{\lambda_{HX}^2}{4} , \nonumber\\ 
 \beta_{\lambda_{HX}}^{(1)} & =&\left(3y_t^2-\frac{9g_1^2}{20}-\frac{9g_2^2}{4}+6\lambda_H+ 3\lambda_X+  6y^2 \right) \lambda_{HX} +2 \lambda_{HX}^2, \nonumber\\ 
 \beta_{\lambda_X}^{(1)} & =&\lambda_{HX}^2+9\lambda_X^2+12y^2 \lambda_X- 12 y^4,\nonumber \\ 
 \beta_{y^2}^{(1)} & =&y^2\left( 9y^2-8 g_3^2 -\frac{18 q^2 g_1^2}{5}\right),\nonumber
\eea
%
where  $g_3$,  $g_2$ and  $g_1=\sqrt{5/3}g_Y$ are the gauge couplings of ${\rm SU(3)}_c $, ${\rm SU(2)_{\it L}}$ and   ${\rm U(1)_{\it Y}}$ respectively and $y_t$ is the top Yukawa.    The RGEs of the massive parameters can be neglected because in this step we are interested in energies several orders of magnitude above the EW scale.

In order to solve the RGEs above we need the values of the involved couplings at some reference energy, which we take to be the top mass $M_t$. As far as the SM couplings are concerned, we will use the central values given in Ref. \cite{Buttazzo:2013uya}. Regarding the new couplings, let us first consider what happens at the mass threshold $M_\psi$. Following  \cite{Casas:1999cd} we adopt the approximation in which the new Yukawa coupling run only above the corresponding mass thresholds; this is  technically implemented by substituting  $y\rightarrow y \theta(\bar{\mu} - M_\psi)$ on the right-hand side of the RGEs. An analogous substitution can be performed to take into account the scalar mass threshold $M_X$: we have to perform the substitutions $\lambda_X\rightarrow \lambda_X \theta(\bar{\mu} - M_X)$ and $\lambda_{HX}\rightarrow \lambda_{HX} \theta(\bar{\mu} - M_X)$ on the right-hand side of the RGEs. The situation is different from the case where the new heavy scalar acquires a VEV: in this case the Higgs quartic coupling gets a tree-level shift  of order $\lambda_{HX}/\lambda_X$  \cite{RandjbarDaemi:2006gf,EliasMiro:2012ay,Dhuria:2015ufo}.  This is the result of integrating out the massive scalar  degree of freedom at  tree-level. The reason why this shift  occurs is  because setting the heavy scalar to zero is not  consistent with the field equations. In the present case $X$ does not develop a VEV and such shift does not occur.

\section{Stability analysis, naturalness and inflation} \label{stab}

Since we use the 1-loop RGEs, we approximate the  Coleman-Weinberg \cite{CW}  effective potential  of this model with its  RG-improved tree-level potential: we substitute the bare couplings in the classical potential with the corresponding running ones. The fact that the $\overline{\rm MS}$ quantities are gauge invariant, as proved in \cite{wil,Buttazzo:2013uya}, guarantees that our results will not be affected by any gauge dependence.


The conditions to guarantee the stability of the EW vacuum at this level of approximation are given in (\ref{stability condition}), where $\lambda_H$ and $\lambda_X$ have to be thought of as the renormalized couplings.

The question of the stability of the EW vacuum has been previously addressed  in other economic extensions of the SM.  For example,   singlet scalars were considered in \cite{Gonderinger:2009jp,EliasMiro:2012ay,Kadastik:2011aa,Salvio:2015cja} and more elaborated possibilities were studied in \cite{Das:2015nwk}.

The sizable value of $y$ needed to reproduce the diphoton excess tends to make $\lambda_X$ negative  at high energies. However, one can compensate this effect via a sizable $\lambda_{HX}$, such that both $\lambda_H$ and $\lambda_X$ remain positive.

\begin{figure}[t]
\begin{center}\includegraphics[scale=0.99]{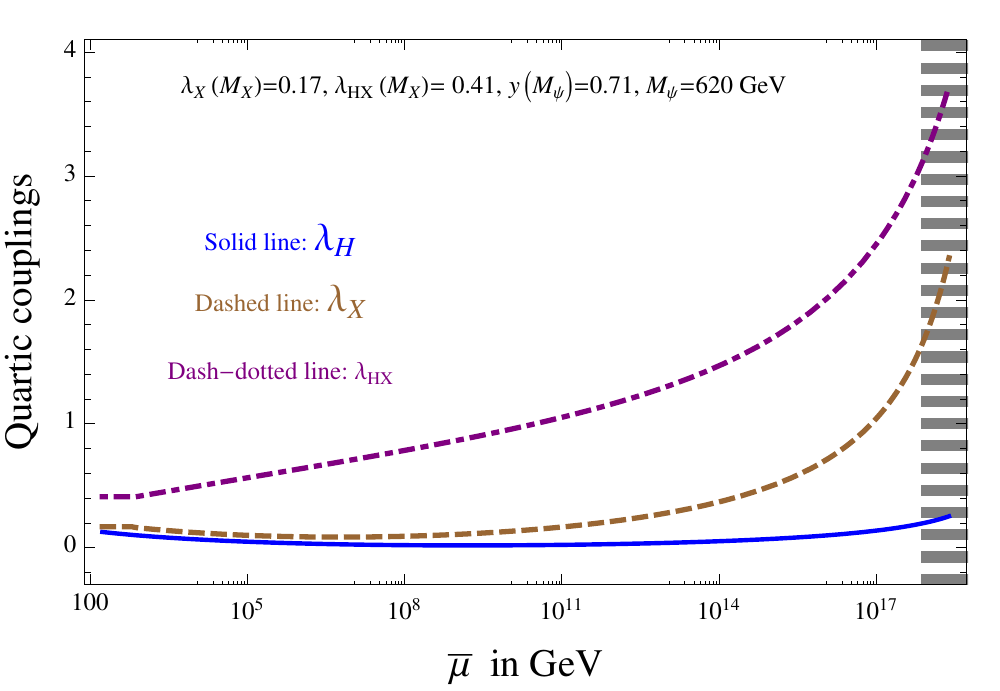} \end{center}
   \caption{{\small \it RG evolution of the quartic couplings $\lambda_H$, $\lambda_X$ and $\lambda_{HX}$ for $M_X\approx 750$ GeV and $q=3/2$.  The stripes on the right indicate the region that is presumably dominated by Planck physics.  } }
\label{RG-evolution}
\end{figure}

\begin{figure}[t]
\begin{center}\includegraphics[scale=0.75]{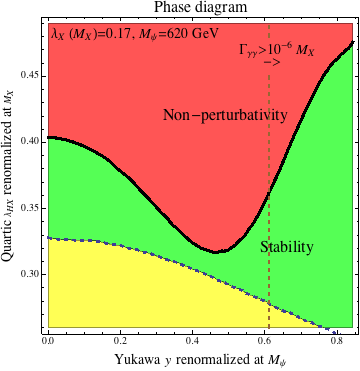} \end{center}
   \caption{{\small \it Phase diagram of the model for $M_X\approx 750$ GeV and $q=3/2$.  We give the stability region (where all couplings are perturbative and the EW vacuum is stable) and the non-perturbativity region. The yellow region below the stability one does not satisfy the stability conditions. The part on the right of the vertical dashed line can account for the diphoton signal.
    } }
\label{phase-diagram}
\end{figure}

In Fig. \ref{RG-evolution}  we show the evolution of the quartic couplings as a function of the renormalization scale. When the parameters are chosen appropriately (e.g. as in that figure), there are no Landau poles in the model, {\it all} couplings remain perturbative and the stability conditions are fulfilled (all couplings are positive) below the Planck scale. Note that the model recently proposed in \cite{Dhuria:2015ufo}, with $q=2$, develops a Landau pole for $g_1$ five orders of magnitude below the Planck scale (and therefore is not perturbative) as it can be shown by solving the RGE of $g_1$ given above. 
 The region with stripes  on the right of  Fig. \ref{RG-evolution} corresponds to the regime where Planck physics is expected to be dominant; the behavior of the curves there is thus presumably unreliable. The  values of the partial decay widths  corresponding to  Fig. \ref{RG-evolution} are 
\be \{\Gamma_{\gamma \gamma} /M_X, \Gamma_{gg}/M_X\} \approx   \{1.3 \times 10^{-6},  \, \, 1.2 \times 10^{-5}\}\label{partial-widths}\ee
and are compatible with the reported signal.  If one increases $M_\psi$ up to the TeV one can still be (although barely) compatible with $\Gamma_{\gamma \gamma}\sim 10^{-6}M_X$,  $\Gamma_{gg}\gtrsim 10^{-6}M_X$ and perturbativity and stability up to the Planck scale by choosing the remaining parameters appropriately: 
\be \{\lambda_X(M_X), \lambda_{HX}(M_X), y(M_\psi), q\} \approx   \{0.2,  \, \, 0.4,  \, \, 0.8,  \, \, 3/2\}.\ee
In Fig. \ref{phase-diagram} we provide a graphic representation of the phase diagram of the model. In the green region all couplings are perturbative and the quartic couplings fulfill the stability conditions, see Eq. (\ref{stability condition}). The non-perturbativity of the red region in Fig. \ref{phase-diagram} is triggered by Landau poles of the quartic couplings, not of the Yukawa $y$.

If such signal will persist, it could be interpreted as a manifestation of the physics needed to stabilize the EW vacuum. Nevertheless, if the signal will turn out to be a statistical fluctuation, we still regard our results interesting because they provide an example of new physics   able to stabilize the EW vacuum without introducing unnaturally large contributions to the Higgs mass $M_h$. In fact in the present model the leading  radiative correction to $M_h^2$ is
\be\delta M_h^2 \approx \frac{C_X \lambda_{HX}}{(4\pi)^2} M_X^2,  \label{deltaMh}
\ee
where $C_X$ is a quantity of order one. In the present case (see e.g. Fig. \ref{RG-evolution}) such correction is not exceeding the order of magnitude of $M_h$. Therefore  a natural EW scale occurs: this  is because all the new particles that have sizable couplings to the Higgs are not too heavy \cite{Farina:2013mla,Salvio:2014soa,Giudice:2014tma}.

We now turn to inflation. Note that $\mathscr{L}_{\rm gravity}$ is non-standard because there are non-minimal couplings between the scalar fields and there is an $R^2$ term. The latter term can be transformed into an additional scalar $z$ and the non-minimal couplings  can be transformed into a non-standard scalar kinetic and potential terms  by going to the so-called Einstein frame: the final scalar-tensor Lagrangian is \cite{Kannike:2015apa}
\be \label{eq:Einstein}
\mathscr{L}_{\rm st} =\sqrt{-g_E} \left(  - \frac{\bp^2}{2} R_E +  \mathscr{L}_{\rm kin} - U
\right),
\ee
where the scalar kinetic and potential terms are 
\bea\label{eq:LKin}
\mathscr{L}_{\rm kin} &=&   \frac{3\bp^2}{z^2}
 \left[(\partial h)^2 + (\partial X)^2+(\partial z)^2\right] \equiv \frac{K_{ij}(\phi) }{2}\partial_\mu \phi^i\partial^\mu \phi^j\\
 U(H,X, z) &=&   \frac{36\bp^4}{z^4}
\bigg[{V(H,X)}+   \frac{1}{16\alpha} \bigg(\frac{z^2}{6} - \xi_H 2 |H|^2-\xi_X X^2 -\bp^2\bigg)^2\bigg] 
\eea
and everything is computed with the metric
\be g^E_{\mu\nu}\equiv g_{\mu\nu}\times  z^2/(6\bar M_{\rm Pl}^2).\ee
Here we have introduced $\{\phi^1, \phi^2, \phi^3\}\equiv \{h,X,z\}$, $K_{ij}$ is   the field metric  and $h$ is the real scalar field associated with the physical Higgs. 
The minimum of $U$ occurs  for $z\approx \sqrt6 \bp$, $X=0$ and $H\approx 0$ (having neglected $v$ at the high inflationary scales).

 The slow-roll conditions can then be  written in a compact form: 
 \be \epsilon \equiv  \frac{  \bar M_{\rm Pl}^2 U_{,i}U^{,i}}{2U^2} \ll 1 \label{1st-slow-roll},\ee 
\be \left|\frac{\eta^{i}_{\,\,\, j} U^{,j}}{U^{,i}}\right|  \ll 1 \quad \mbox{($i$  not summed), }\quad \mbox{where}\quad \eta^{i}_{\,\,\, j}\equiv \frac{\bar M_{\rm Pl}^2 U^{;i}_{\,\,\, ;j}}{U}  \label{2nd-slow-roll} \ee
and we raise the indices with  the inverse field metric $K^{ij}$;  also,
 $U_{,i} \equiv \partial U/\partial \phi^i$ and $U_{;i}$ is the covariant derivative built with $K_{ij}$. When these conditions are fulfilled the field equations are 
 \be  \dot \phi^i\approx -\frac{U^{,i}}{3H},\qquad H^2 \approx \frac{U}{3  \bar M_{\rm Pl}^2}, \label{slow-roll-eq}\ee
where a dot denotes the derivative w.r.t. the cosmic time $t$; also $H\equiv \dot a /a$ and $a$ is the Robertson-Walker scale factor.

\begin{figure}[t]
\begin{center}\includegraphics[scale=0.9]{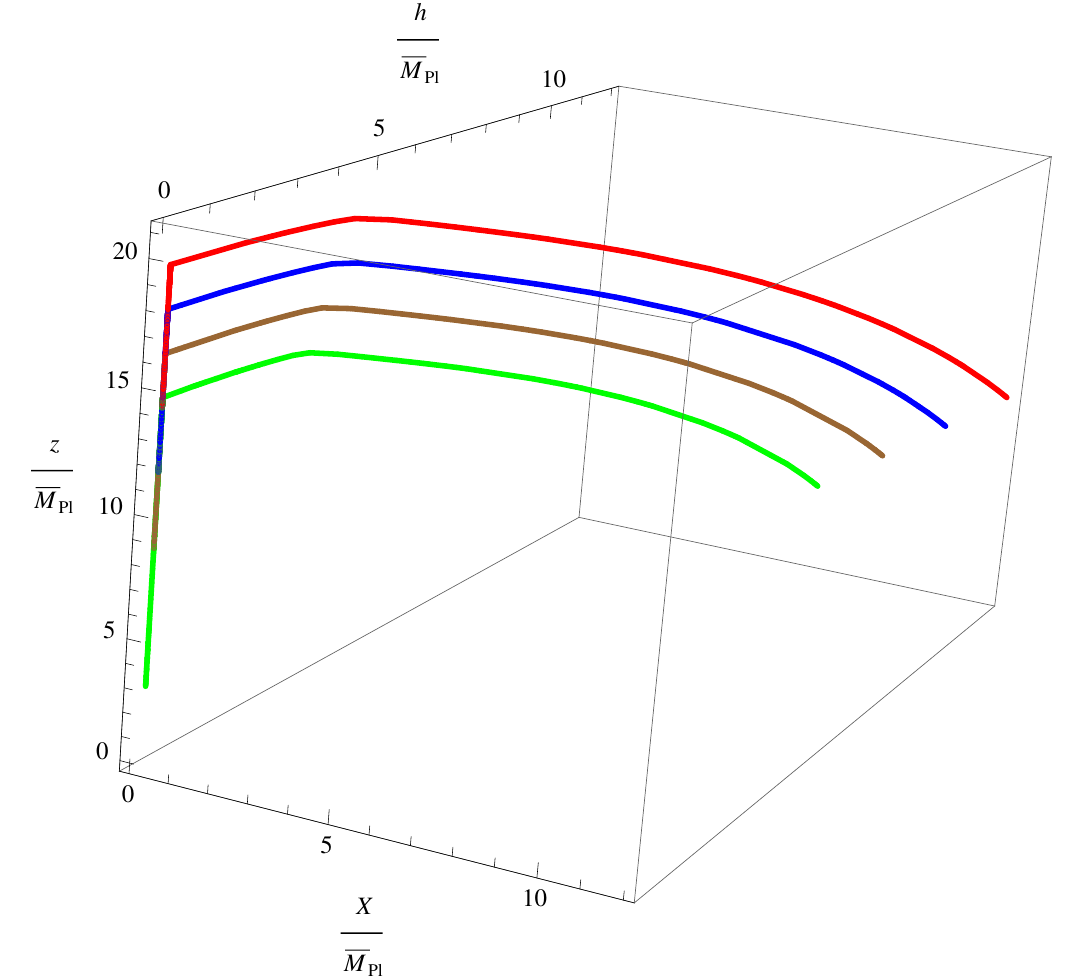} \end{center}
   \caption{{\small \it Path in scalar field space obtained by solving the field equations in (\ref{slow-roll-eq}). Here we set $\xi_H=\xi_X=-1/6$, the value of $\alpha$ is fixed by the observed amplitude of scalar fluctuations (for a number of e-folds of about 60) and the remaining parameters are chosen like in Fig. \ref{RG-evolution}.} }
\label{field-path}
\end{figure}

The sizable quartic couplings in Fig. \ref{RG-evolution} at the inflationary scales suggest that inflation is driven by $z$ rather than the Higgs  or $X$.  Indeed, sizable quartic couplings correspond to field directions with steep potential. In Fig. \ref{field-path} we show that this is indeed the case: even if we start from initial values of $h$ and $X$ larger than $z$ the dynamics converge to the line $h=X=0$ and only there the slow-roll conditions in (\ref{1st-slow-roll}) and (\ref{2nd-slow-roll}) are satisfied. The inflationary predictions are therefore those of Starobinsky's model \cite{Starobinsky}, in agreement with current observations. 
 
 Even if one set initially $\alpha=0$, Higgs inflation \cite{Bezrukov:2007ep} or $X$-inflation would generically require large values of $\xi$, given the present bound on the tensor to scalar ratio $r\lesssim0.1$ \cite{Ade:2015tva} and such large non-minimal couplings  would source a large value of $\alpha$ modulo fine-tuning \cite{Salvio:2015kka}.  The same conclusion is reached even if one does not introduce $X$ and the only scalar fields are $h$ and $z$.

\section{Conclusions}	\label{Conclusions}

In this paper, we considered an extra CP-odd scalar $X$ and an extra vector-like fermion $\Psi$ carrying color and electric charge: this is the simplest model that can explain the excess in the diphoton channel reported by ATLAS and CMS  at $750$~GeV. 

Our 
computations, valid up to one-loop in the RGEs, suggest that it is indeed possible to explain the excess in the diphoton channel, curing at the same time the SM instability  of the EW vacuum and keeping {\it all} couplings perturbative up to the Planck scale. An example is provided in Fig. \ref{RG-evolution}, which uses a parameter choice leading to the partial decay rates of $X$ in Eq. (\ref{partial-widths}). The phase diagram with the regions of stability (which corresponds to perturbative couplings and stable EW vacuum), non-perturbativity and unstable EW vacuum are given in Fig. \ref{phase-diagram}. There, the part that can account for the diphoton signal is also provided.

Besides the excess in the diphoton channel, we believe that these results are interesting for another reason, independently on whether or not the ATLAS and CMS excess at 750 GeV will be confirmed by future analysis: they provide a SM extension that can stabilize the EW vacuum without invoking an unnaturally large contribution to the Higgs mass. Indeed, in the present model $M_h$ is natural as  the largest correction to it, given by Eq. (\ref{deltaMh}),  is never $\gg M_h$.

Moreover, we studied the classical dynamics of the model during inflation by considering the most general Lagrangian with operators of dimension up to 4 and including the new scalar $X$.
Although,  both $X$ and/or the SM Higgs can potentially inflate the early universe 
and possibly explain the current CMB data, the sizable values of the quartic couplings needed to stabilize the EW vacuum forces inflation to be driven by the effective scalar $z$, which emerges from the $R^2$ operator: we 
found that during inflation, the classical slow-roll trajectory is dominated by $z$ and both $X$ and $h$ roll down quickly within few Hubble times due to the sizable quartic couplings. Moreover, another  reason favors $z$-inflation (also known as Starobinsky inflation \cite{Starobinsky}). Large non-minimal couplings $\xi_H$ and $\xi_X$ to gravity are generically required in order
to satisfy the right amplitude of scalar perturbations in the CMB and the current constraint on the tensor-to-scalar ratio; indeed, as analyzed before in Ref. \cite{Salvio:2015kka}, a large non-minimal coupling to gravity would yield higher order curvature corrections (e.g. $R^2$ terms) modulo fine-tuning.

\vspace{1.5cm}
\noindent {\bf Acknowledgments.} We thank  Bhupal Dev,  Jong Soo Kim, Jes\'us M. Moreno,  Krzysztof Rolbiecki for useful   discussions.  This work has been supported by the Spanish Ministry of Economy and Competitiveness under grant FPA2012-32828,  Consolider-CPAN (CSD2007-00042), the grant  SEV-2012-0249 of the ``Centro de Excelencia Severo Ochoa'' Programme, the grant  HEPHACOS-S2009/ESP1473 from the C.A. de Madrid and the ERC grant NEO-NAT.

\vspace{0.9cm}
\noindent {\bf Note added.} While we were finalizing this paper another work \cite{Zhang:2015uuo} that addresses the EW vacuum instability appeared in the arXiv. Moreover, other articles, which are related to the present analysis, appeared \cite{Knapen:2015dap}. After our article was posted on the arXiv, Ref. \cite{Dhuria:2015ufo} was updated admitting that the $q=2$ case has a Landau pole much below the Planck scale.

\vspace{0.9cm}

 \begin{multicols}{2}

\end{multicols}

\end{document}